\documentclass[]{spie}  

\usepackage[]{graphicx}

\title{The X-/Gamma-ray camera ECLAIRs for the Gammay-ray burst mission SVOM} 

\author{Godet O.\supit{a,b}, Nasser G.\supit{a,b}, Atteia J.-L.\supit{a,b},
  Cordier B.\supit{c}, Mandrou P.\supit{a,b}, Barret D.\supit{a,b}, Triou
  H.\supit{c}, Pons R.\supit{a,b}, Amoros C.\supit{a,b}, Bordon S.\supit{a,b},
  Gevin O.\supit{c}, Gonzalez F.\supit{e}, G\"otz, D.\supit{c},
  A. Gros\supit{c}, Houret B.\supit{a,b}, Lachaud C.\supit{d}, Lacombe
  K.\supit{a,b}, Marty W.\supit{a,b}, Mercier K.\supit{e}, Rambaud
  D.\supit{a,b}, Ramon P.\supit{a,b}, Rouaix G.\supit{a,b}, Schanne
  S.\supit{c}, Waegebaert V.\supit{a,b} \skiplinehalf \supit{a} Universit\'e
  de Toulouse, OMP-UPS, Institut de Recherche en Astrophysique et
  Plan\'etologie (IRAP), Toulouse, France; \\ \supit{b} CNRS, IRAP, 9 avenue
  du colonel Roche, 31028 Toulouse Cedex 4, France;\\ \supit{c} CEA Saclay,
  DSM/IRFU/service d'Astrophysique, 91191, Gif-sur-Yvette, France;\\ \supit{d}
  AstroParticule et Cosmologie (APC), CNRS-UMR 7164, Université de Paris 7
  Denis Diderot, 75205, Paris, France;\\ \supit{e} CNES, 18 Av. Edouard Belin,
  31401 Toulouse cedex 9, France;\\ }

\authorinfo{Further author information: (Send correspondence to
  O. Godet)\\Olivier Godet: E-mail: ogodet@irap.omp.eu, Telephone: 33 561 55
  7536}

\begin{document} 
  \maketitle 

\begin{abstract}

We present ECLAIRs, the Gamma-ray burst (GRB) trigger camera to fly on-board
the Chinese-French mission SVOM. ECLAIRs is a wide-field ($\sim 2$\,sr) coded
mask camera with a mask transparency of 40\% and a 1024 $\mathrm{cm}^2$
detection plane coupled to a data processing unit, so-called UGTS, which is in
charge of locating GRBs in near real time thanks to image and rate
triggers. We present the instrument science requirements and how the design of
ECLAIRs has been optimized to increase its sensitivity to
high-redshift GRBs and low-luminosity GRBs in the local Universe, by having a
low-energy threshold of 4 keV. The total spectral coverage ranges from 4 to
150 keV. ECLAIRs is expected to detect $\sim 200$ GRBs of all types during the
nominal 3 year mission lifetime.

To reach a 4 keV low-energy threshold, the ECLAIRs detection plane is paved
with 6400 $4\times 4~\mathrm{mm}^2$ and 1 mm-thick Schottky CdTe
detectors. The detectors are grouped by 32, in 8x4 matrices read by a
low-noise ASIC, forming elementary modules called XRDPIX. In this paper, we
also present our current efforts to investigate the performance of these
modules with their front-end electronics when illuminated by charged particles
and/or photons using radioactive sources. All measurements are made in
different instrument configurations in vacuum and with a nominal in-flight
detector temperature of $-20^\circ$C. This work will enable us to choose the
in-flight configuration that will make the best compromise between the science
performance and the in-flight operability of ECLAIRs. We will show some
highlights of this work.

\end{abstract}



\section{The SVOM mission}
\label{sec:intro}  

\subsection{Catching the most powerful transient events in
  the Universe}

Amongst the most powerful transient events known, Gamma-ray bursts (GRBs)
appear to be the most extreme ones with isotropic radiated emission up to
$\sim 10^{54}$ erg. They are highly transient and powerful cosmological
explosions appearing randomly in the sky and characterized by very short (from
a few milli-seconds to hundreds of seconds) prompt X-/Gamma-ray flashes
followed by a multi-wavelength long-lasting afterglow emission (from days to
months in some cases). They are often divided into two categories depending on
their duration: short GRBs with duration typically less than 2\,s and long
GRBs with duration typically larger than 10\,s
(e.g. [\citenum{Kouveliotou93}]), even if this classification could sometimes
be misleading (e.g. [\citenum{Zhang07}]). They are thought to be associated
with the catastrophic formation of stellar mass black holes coupled with the
launch of ultra-relativistic jets when some massive stars with masses larger
than $10~M_\odot$ collapse (for long GRBs e.g. [\citenum{Woosley99}]) or when
two neutron stars merge (e.g. [\citenum{Eichler89}]) for short GRBs.

GRB prompt and afterglow emission convey broadband information on the physics
of the GRB phenomenon itself (e.g. particle acceleration in the jet and the
radiation mechanisms), but also on the nature of the GRB progenitors with the
detection of supernovae associated with long GRBs (e.g. [\citenum{Galama98,
    Campana06}]) and the recent detection of kilonovae associated with some
short GRBs (e.g. [\citenum{Tanvir13, Berger13}]). Progress on these topics has
been made possible with the launch of the {\it Swift} mission
[\citenum{Gehrels04}] in 2004, in particular on the afterglow and the nature
of the progenitors of short GRBs (e.g. [\citenum{Gehrels05, B05, Church11,
    Berger11, K12, Tanvir13}]) as well as the characterization of the early
afterglow (see e.g. [\citenum{Tagliaferri05, Nousek06, OB06, Chincarini07,
    Kann10, Godet11}]).  However, there are still some puzzling questions that
have to be answered on the prompt and afterglow emission as summarized in
[e.g. \citenum{Zhang07, Godet11, BZ12}].  To be able to answer them, well
characterized (spectrally and temporally) multi-wavelength datasets of both
prompt and afterglow emission of GRBs with measured redshifts are needed from
future GRB missions along with some essential pieces of information to be
found from particle and gravitational wave messengers.  The detection of shock
breakouts associated with normal SNe [e.g. \citenum{Soderberg08}] and
low-luminosity GRBs [e.g. \citenum{Campana06}] offers a unique way to probe
the connection between GRBs and classical supernovae, and how the core of a
massive star collapses.

GRBs also convey a lot of excitement in cosmology and fundamental physics
(e.g. ultra high-energy cosmic rays, high energy neutrinos and gravitational
waves). Indeed, being very bright and located at cosmological distances -- the
farthest known GRB has a redshift of 8.2 ([\citenum{tanvir09}], see also
[\citenum{Cucchiara11}] with a photometric redshift at 9.4 measured for GRB
090429B] -- they pinpoint their host galaxy offering a way to study the first
  galaxies and maybe the first stars of the Universe, the so-called population
  III stars. They could be used to put constraints on a poorly known phase,
  the so-called re-ionization phase, when the matter of the Universe is
  thought to have been re-ionized by UV/X-ray photons from the first
  structures (stars and quasars), to track the evolution of baryons with
  redshift by studying the foreground material illuminated along their line of
  sight. If some properties of GRBs can be standardized as it is done for
  type-Ia supernovae, GRBs will offer a unique way to study the expansion of
  the Universe beyond redshift 2.  Coalescence of two neutron stars
  thought to be the progenitors of some short GRBs are ideal sources to detect
  gravitational waves. Therefore, short GRBs offer an ideal way to pinpoint
  such systems on the sky for ground-based GW facilities such as aLIGO/aVirgo
  [\citenum{Corsi12}]. The recent and future detections of kilo-novae
  associated with short GRBs also offer a way to study the nucleosynthesis of
  r-process elements and how they enrich the interstellar/intergalactic medium
  [e.g. \citenum{Tanvir13, Berger13}].

The newly discovered class of ultra-long GRBs lasting several hundreds of
minutes [e.g. \citenum{Gendre13, Levan14}] might indicate the existence of
either a new class of star progenitor with much larger radii (e.g. blue
giants) or another type of objects as the disruption of a white dwarf by an
intermediate mass BH (IMBH -- [\citenum{Levan14, S13, MacLeod14}]). If it is
the case, this offers very exciting prospects to detect strong GW signals
[e.g. \citenum{Sesana08}] and the long sought IMBH class that might be the
building bricks to form supermassive black holes (SMBH) and that stay rather
elusive (the only strong IMBH candidate being the ultra-luminous X-ray source
HLX-1 in ESO 243-49 -- [\citenum{Farrell09}]).

Tidal Disruption Events (TDEs -- \citenum{Komossa12}) are also an interesting
class of energetic transient events as {\it Swift} showed by recently catching
in real time a star being disrupted by a supermassive BH
[\citenum{Burrows11}]. TDEs offer the possibility to probe normally quiet
supermassive BHs and hence to better constrain the fraction of occupation of
BHs in the core of galaxies. The number of detected TDEs is likely to strongly
increase in future years with monitoring of the transient sky in optical with
the Large Synoptic Survey Telescope (LSST) and in radio with LOFAR
(Low-Frequency Array for radio astronomy) and SKA (Square Kilometre Array).

\begin{figure}
\begin{center}
\begin{tabular}{c}
\includegraphics[height=8.5cm]{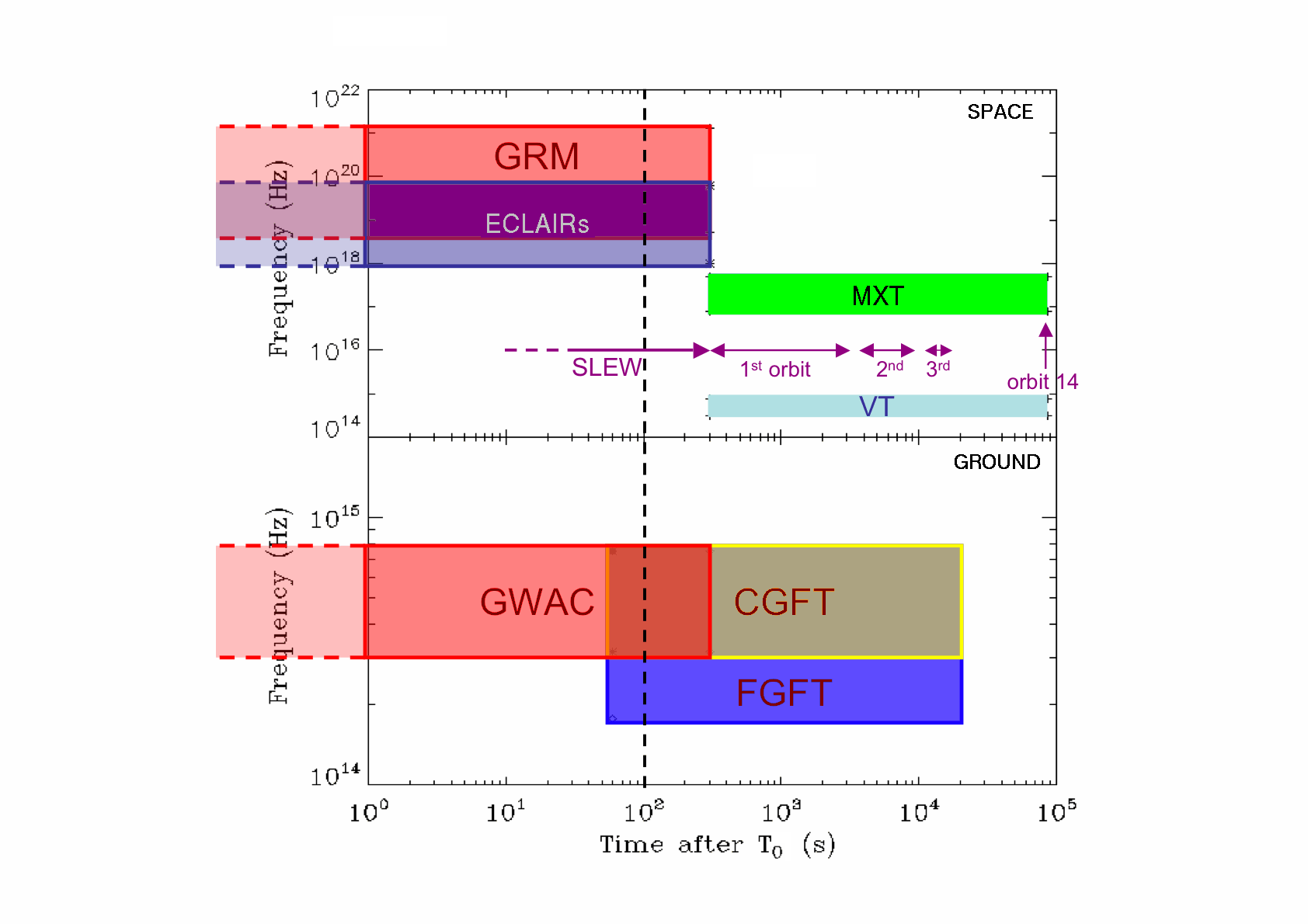}
\end{tabular}
\end{center}
\caption[example]{\label{fig:spectralcoverage} Broadband coverage from the
  space and ground instruments of SVOM as a function of time after the ECLAIRs
  GRB trigger $T_0$.}    
\end{figure}

Given this exciting and very promising science context for the future, the
Chinese National Space Agency and the French Space Agency (CNES) have decided
to jointly implement a mini-satellite mission dedicated to the study of
Gamma-ray bursts and other powerful transient events for a launch around
2020. The mission SVOM (Space borne Variable Object Monitor) will operate \`a
la {\it Swift} {\it i.e.} a platform consisting of a multi-wavelength payload
coupled to rapid autonomous slewing capability [\citenum{Godet12} -- see
  Mercier et al. - paper 9144-73 this conference]. The science payload will
consist of 2 wide-field cameras: the GRB trigger coded-mask camera ECLAIRs
[\citenum{Triou09}] and the Gamma-Ray Monitor [\citenum{Dong09}] (see Wu et
al.: paper 9144-195 this conference) as well as two narrow-field instruments,
the Micro X-ray channel plate Telescope (MXT -- [\citenum{perinati12}]; see
also G\"otz et al. paper 9144-74 this conference) working in the 0.3-10 keV
band and the Visible Telescope (VT -- [\citenum{Wang10, Wu12}]) covering the B
\& R visible bands. The development of the ECLAIRs and MXT instruments is led
by the CNES in collaboration with several French and European astrophysics
institutes. The IRAP is Principal Investigator of ECLAIRs, while the CEA is PI
of the MXT. In addition to the space payload, SVOM will have a dedicated
network of robotic narrow-field Ground Follow-up Telescopes (GFT) and Ground
Wide Angle optical (V band) Cameras (GWAC) placed along the SVOM orbit in
China and in Mexico. The French GFT (PI: IRAP) will have near infrared
capability (J \& H bands) and maybe low resolution spectroscopy capability in
addition to the use of the B, V, R, I filters, while the Chinese GFT will
cover optical wavelengths. The GWAC camera led by China will be able to
monitor a fraction of the ECLAIRs field of view (FoV) in order to search for
optical prompt emission as observed for the naked eyed burst GRB\,080319B
[\citenum{racusin08}] and other optical transients.
Figure~\ref{fig:spectralcoverage} gives an overview of the SVOM capability in
terms of its temporal response as well as its spectral coverage from ground to
space.

\subsection{Observing strategy}
\label{sec:strategy2}

\begin{figure}
\begin{center}
\begin{tabular}{c}
\includegraphics[height=6cm]{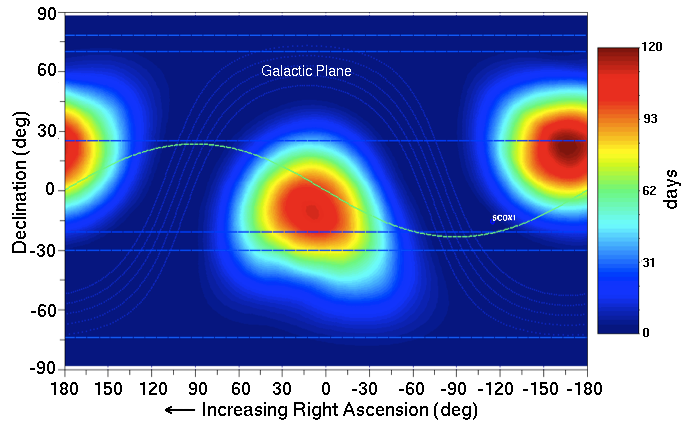}
\end{tabular}
\end{center}
\caption[] { \label{fig:exposure} Sky map showing the pointing directions of
  the GRB trigger camera ECLAIRs. The SVOM orbit will be nearly anti-Sun and
  will avoid having the Galactic plane \& Sco X-1 in the field of view of
  ECLAIRs. The pointing color scale corresponds to the number of days during
  which a sky position will be in the field of view of ECLAIRs.}
\end{figure}

SVOM will be operated in a Low Earth Orbit with an altitude of $625-650$ km
and an inclination around $30^\circ$. In order to facilitate the observations
of SVOM GRBs by the largest ground based telescopes, and hence to increase the
fraction of GRBs with measured redshift to be larger than $ 50\%$ (against
nearly a third for the {\it Swift} mission from October 2004 to June
2014\,\footnote{The {\it Swift}-BAT has detected 878 GRBs until June 2014, of
  which 278 GRBs had a measured redshift.}), SVOM will implement a nearly
anti-solar pointing strategy with general avoidance of the Galactic center,
Sco X-1 and the Moon in the ECLAIRs FoV in order not to reduce
too much the ECLAIRs sensitivity to GRBs (see Fig.~\ref{fig:exposure}). In
addition, for GRBs detected in the Galactic plane it will be harder to search
for an optical counterpart due to dust extinction in the Galactic plane.

SVOM will be operated \`a la {\it Swift} in order to catch GRBs: ECLAIRs will
scan specific regions of the sky compliant with the spacecraft attitude law
(see Fig.~\ref{fig:exposure}) and the GRM will observe the same FoV. For $\sim
20$\% of the detected GRBs, the GWAC will be able to search for optical prompt
emission on the ground. At $t=T_0$, ECLAIRS triggers on a burst, computes a
position, sends it to the ground in near real time (less than 1 minute) via a
network of VHF stations distributed along the SVOM orbit and requests a slew
to the spacecraft in order to repoint the satellite so that the ECLAIRs GRB
position is brought in the FoV of the MXT and VT in less than 5 min after
$T_0$. Once possible, one of the two GFTs on the ground will start to observe
the ECLAIRs error circle 1 min after $T_0$ in order to detect an optical
afterglow counterpart and then to refine the positions. Each time a refined
position is found by any of the SVOM instruments, an alert message will be
promptly distributed to the world community.

\subsection{Mission requirements} 
\label{sec:requirement}

SVOM was conceived as a mini-satellite mission enabling further understanding
of the GRB phenomenon and its related physics (and other energetic transient
events) by providing a well characterized (spectrally and temporally)
multi-wavelength dataset. This led us to define the science requirements
listed below:
\begin{enumerate}
\item To enable the detection of all known types of GRBs, and particularly
high-z GRBs for cosmological studies and low-z, sub-luminous GRBs to study in
detail the connection with supernovae;

\item To quickly identify the rapidly decaying flux afterglows in X-rays and in optical;

\item To provide fast, reliable and accurate GRB positions (with accuracy less
  than 1 arcminute in X-rays and sub-arsec in optical) in order to maximize
  the probability of measuring a redshift and to perform cosmological studies
  with the large on-ground facilities;

\item To measure the temporal properties and the broadband spectral shape of
the prompt emission (from visible to MeV) in order to measure all the spectral
parameters of the prompt emission, in particular the $E_{peak}$ parameter;

\item To quickly provide some GRB positions and parameters to encourage
  follow-up observations.
\end{enumerate}

To implement such mission requirements has a profound impact on the design of
the multi-wavelength payload as well as on the choice of the mission observing
strategy as discussed in Section~\ref{sec:strategy2}.  In this paper, we focus
on the GRB trigger camera ECLAIRs.  First, we present the expected ECLAIRs
performance (Section~\ref{sec:perfo}) to show that SVOM will be a strong
contributor in the GRB field after 2020 and will also offer a good instrument
sensitivity from optical to keV to perform non GRB science
(Section~\ref{sec:nongrb}).  In Section~\ref{sec:design}, we present how the
mission requirements are broken down into technical specifications for the
design of the ECLAIRs camera, in particular for the detection plane.  We will
also describe our effort to characterize the performance of the elementary
detector matrices that will pave the ECLAIRs detection plane (XRDPIX) and show
some highlights of this work in
Section~\ref{sec:calibration}. Section~\ref{sec:conclusion} is devoted to the
conclusion.

\section{Science performance}
\label{sec:perfo}

In this Section, we give an overview of the main expected performance of the
ECLAIRs camera for GRB and non-GRB science. 

\subsection{GRB localization accuracy}
\label{sec:localisation}

Given the design constraints on ECLAIRs (see Section~\ref{sec:design}), a work
of optimization of the mask pattern has been made with several objectives: i)
to define a mask pattern with acceptable thermo-mechanical stresses and hence
avoiding too large deformations that will introduce systematics in the image
reconstruction; ii) to have a reasonably good positioning accuracy to localize
GRBs (and other transient events) as well as to maximize the GRB sensitivity
of ECLAIRs (Gros et al. in preparation). To do so, we studied the performance
of masks with quasi-random patterns with different $\frac{m}{d}$ ratios with
$m$ and $d$ being the linear size of a smallest mask hole and a CdTe pixel on
the detection plane, respectively. From this work, the current baseline mask
with $\frac{m}{d}=2.6$ is expected to provide a positioning accuracy of 13.9
arcminutes ($90\%$ confidence level) for faint GRBs (the lower detection SNR
threshold is currently set at 7) and a few arcminutes for bright ones (see the
left panel in Fig~\ref{fig:sensitivity}). This value does not include the
systematic errors. This mask also provides a gain in the GRB detection
sensitivity of the order of 30\% when compared to a mask with an initially (or
previously) proposed value of $\frac{m}{d}=1.2$.  The design of the MXT \& VT
FoVs took into account the ECLAIRs positioning accuracy for faint GRBs in
order to ensure the necessary refinement of the primary GRB position to a few
dozens of arc-seconds with the MXT [see G\"otz et al. - paper 9144-74 this
  conference] and sub-arcsecond accuracy with the VT.

\subsection{Sensitivity to GRBs}
\label{sec:sensitivity}

\begin{figure}
\begin{center}
\begin{tabular}{cc}
\includegraphics[height=6.cm]{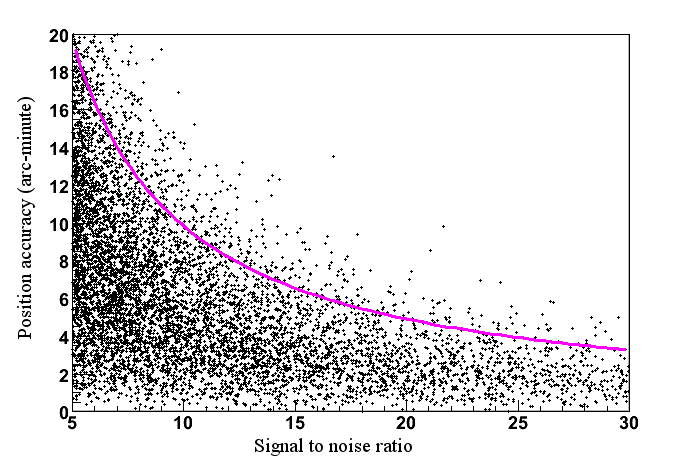} &\hspace{-1cm}
\includegraphics[height=6.cm]{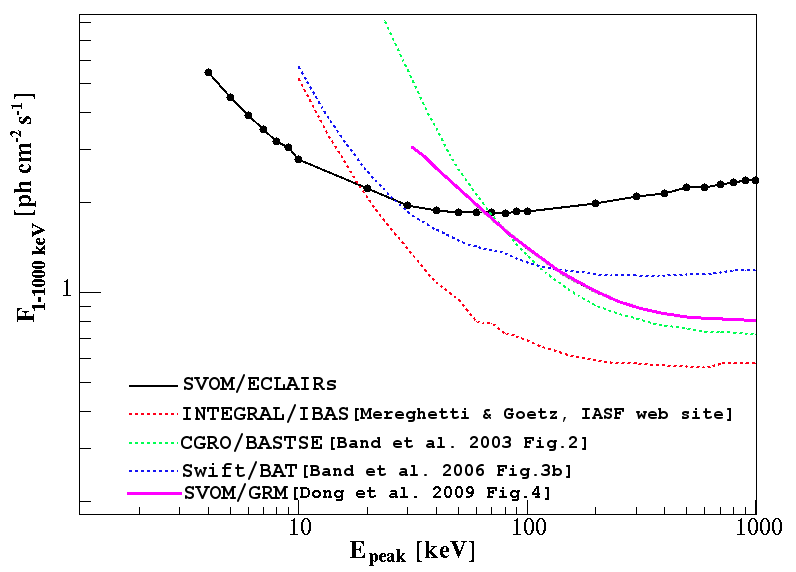} \\
\end{tabular}
\end{center}
\caption[]{\label{fig:sensitivity} Left -- Estimation of the ECLAIRs
  localization accuracy as a function of the estimated GRB signal-to-noise. We
  simulated $10^5$ GRBs with various properties within the ECLAIRs FoV in
  order to compute the GRB position accuracy using the ECLAIRs trigger
  algorithm simulator.  The solid curve corresponds to the 90\% confidence
  level error radius. Right -- Simulated 1-1000 keV on-axis peak sensitivity
  of the GRB trigger camera ECLAIRs compared to other instruments as a
  function of the $E_{peak}$ parameter. We assumed a signal-to-noise ratio of
  $5.5~\sigma$, a GRB Band function with $\alpha = -1$ \& $\beta = -3$ and an
  ECLAIRs imaging band from 4 to 50 keV. The SVOM/GRM sensitivity comes from
  [\citenum{Dong09}].}
\end{figure}

Fig.~\ref{fig:sensitivity} (right panel) compares the $5.5~\sigma$ peak
sensitivity for an on-axis GRB computed for ECLAIRs with other wide-field
instruments: {\it Swift}-BAT, INTEGRAL/IBAS, CGRO/BATSE \& SVOM/GRM as a
function of the GRB hardness. From this Figure, it is clear that ECLAIRs,
thanks to a low energy threshold of 4 keV, will be as sensitive as the {\it
  Swift}-BAT for GRBs with $E_{peak}$-values less than 20 keV (i.e. X-ray
Flashes and potential high redshift GRBs).

\begin{figure}
\begin{center}
\begin{tabular}{c}
\includegraphics[height=5.5cm]{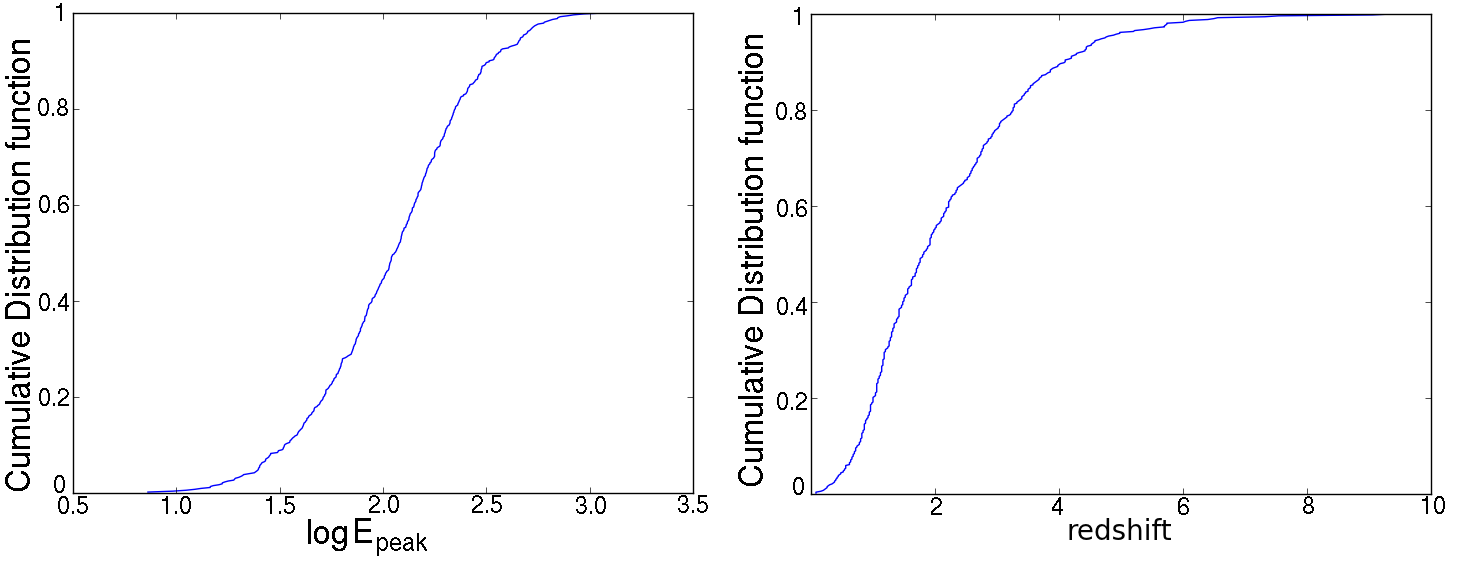}
\end{tabular}
\end{center}
\caption[]{\label{fig:cdf} Simulated cumulative
  distribution function of GRBs detected by ECLAIRs as a function of
  $E_{peak}$ ({\it left}) and redshift ({\it right}).}
\end{figure}

Using the fraction of observed sky per year (that includes the ECLAIRs FoV,
the Earth and SAA crossing), the expected level of the background in the
ECLAIRs energy range [e.g. \citenum{Godet09}] and the $\log N - \log S$
function for long GRBs from [\citenum{Butler10}], ECLAIRs is expected to be
able to detect at least $\sim 180$ long GRBs over the 3 yr lifetime of the
mission. We assumed here that a GRB is detected by ECLAIRs when the
signal-to-noise ratio is more than 6. At least 10\% should be added to take
the short GRB population into account, giving a total of $\sim 200$ GRBs. An
intensive work has been realized in order to refine this estimate using the
ECLAIRs GRB trigger simulator, a physically-based synthetic GRB population \&
the estimated in-orbit instrument background and taking into account the
constraints induced by the SVOM pointing strategy (See
Section~\ref{sec:strategy2}), in particular the times when the Earth limb is
within the ECLAIRs FoV. This fine work finds similar estimate for the ECLAIRs
GRB detection rate.  Figure~\ref{fig:cdf} shows some predictions for the
properties of the GRBs detected by ECLAIRs as a function of redshift and GRB
hardness. 4\% of ECLAIRs GRBs are expected to be high redshift GRBs with $z >
5$; which corresponds to a detection rate of nearly 3 high-z GRBs per
year. This rate is comparable to the one of {\it Swift}-BAT. Most of ECLAIRs
GRBs are also predicted to have $E_{peak}$ values less than 500 keV,
indicating that for most SVOM GRBs it will be possible to measure the spectral
parameters of the prompt emission using ECLAIRs and GRM data (see also
Fig.~\ref{fig:sensitivity}).

\subsection{Non GRB science} 
\label{sec:nongrb}

As for previous GRB missions, the SVOM observing time dedicated to GRB science
will only represent a small fraction of the total observing time. In order to
maximize the mission science return and to take advantage of the broad band
coverage of the SVOM science payload, the SVOM consortium will implement a
general non-GRB science program that will consists of: Target of Opportunities
(ToOs) for unique objects \& a Guest Observer program opened to the whole
community via a contact with one of the SVOM co-investigators, as well as key
programs (e.g. monitoring of transient objects within the Galactic Center;
follow-ups of some of the LSST, LOFAR, SKA or CTA transient
events). Discussions about the key programs are still on-going within the SVOM
consortium. The time devoted to the non-GRB science program is also expected
to increase during the mission lifetime.

\begin{figure}
\begin{center}
\begin{tabular}{c}
\includegraphics[height=5cm]{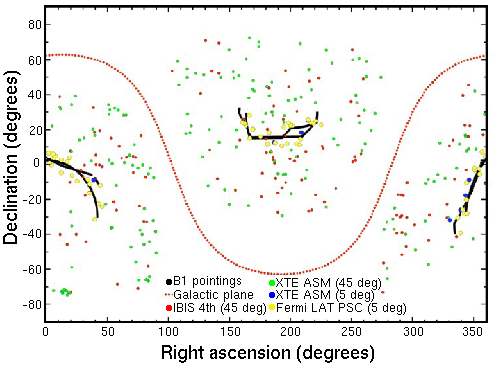}
\end{tabular}
\end{center}
\caption[example] { \label{fig:target} Sky map showing potential targets
  within the ECLAIRs FoV ($\pm 45^\circ$ to the B1 law).}
\end{figure}

\begin{figure}
\begin{center}
\begin{tabular}{c}
\includegraphics[height=5.cm]{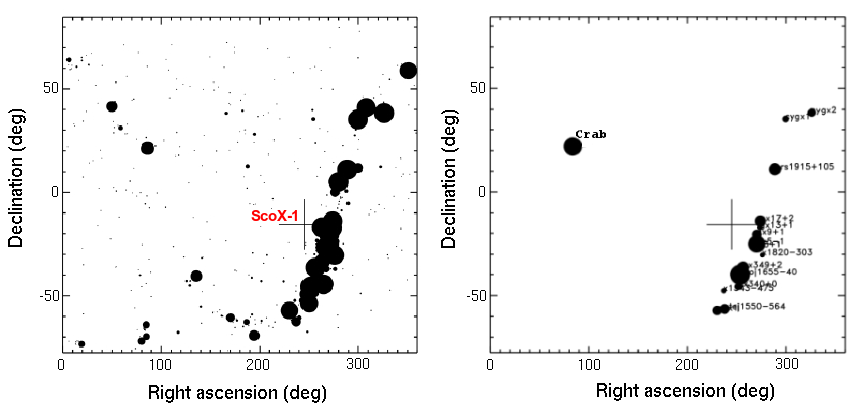}
\end{tabular}
\end{center}
\caption[example]{\label{fig:survey} Left: The X-ray sky as seen by the RXTE
  All-Sky Monitor. Most of the X-ray sources are persistent. The size of the
  dots is proportional to the brightness of the X-ray sources. The cross
  indicates the location of Sco\,X-1. The sources outside the Galactic plane
  are mostly AGN. Right: The X-ray sky above 50 mCrab that roughly corresponds
  to the ECLAIRs sensitivity in just one orbit.}
\end{figure}

Here we outline potential targets that could be observed with ECLAIRs given
the SVOM pointing strategy and the instrument sensitivity ($\sim
50~\mathrm{mCrab}$ in an orbit in the 4-150 keV energy band). For ECLAIRs, any
sources located at $\pm 45^\circ$ from the pointing law could be observed (see
Fig.~\ref{fig:exposure}). Figures~\ref{fig:target} \& \ref{fig:survey} shows
an example of potential targets accessible to ECLAIRs.

\section{Current design of the ECLAIRs GRB trigger camera} 
\label{sec:design}

\begin{figure}
\begin{center}
\begin{tabular}{cc}
\includegraphics[height=5.7cm]{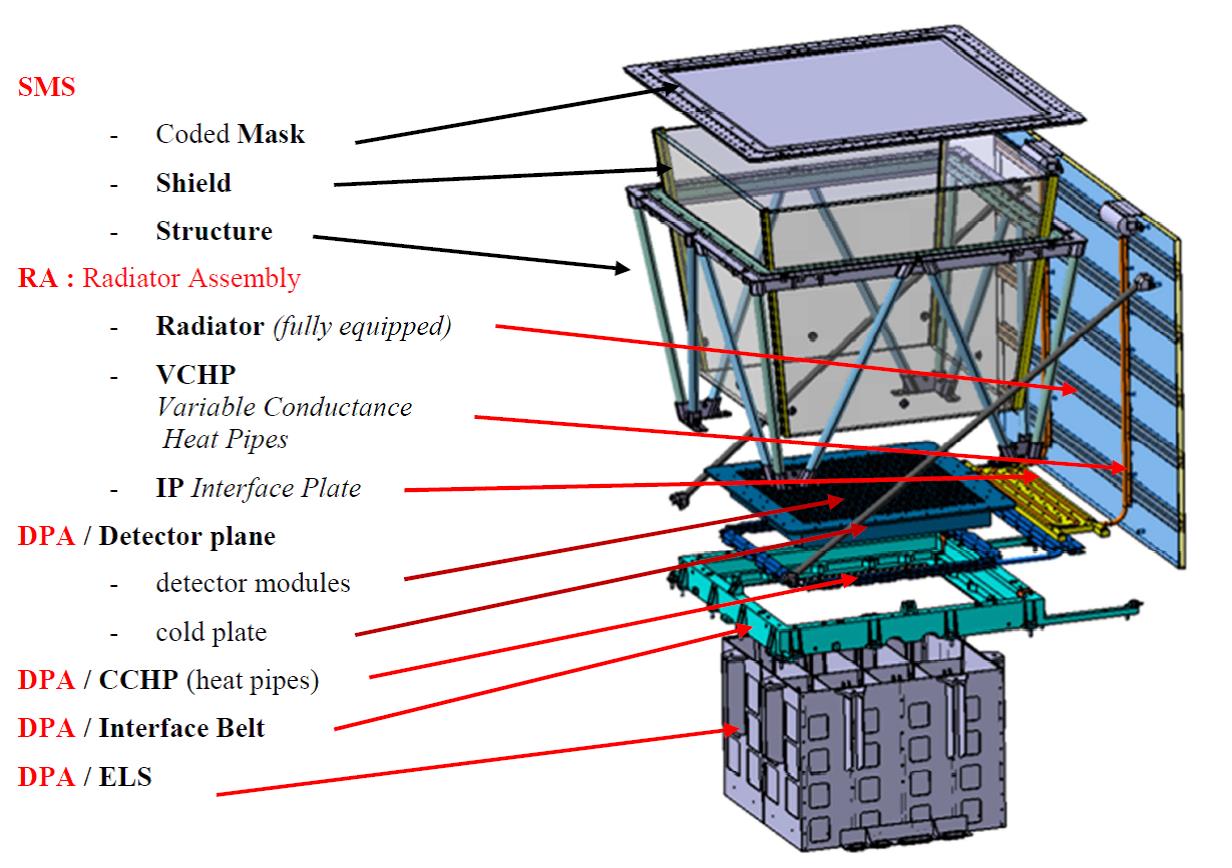} &\hspace{-0.8cm}
\includegraphics[height=5.cm]{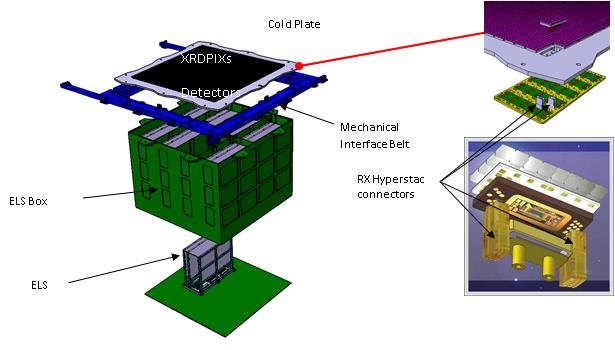}\\
\includegraphics[height=4.cm]{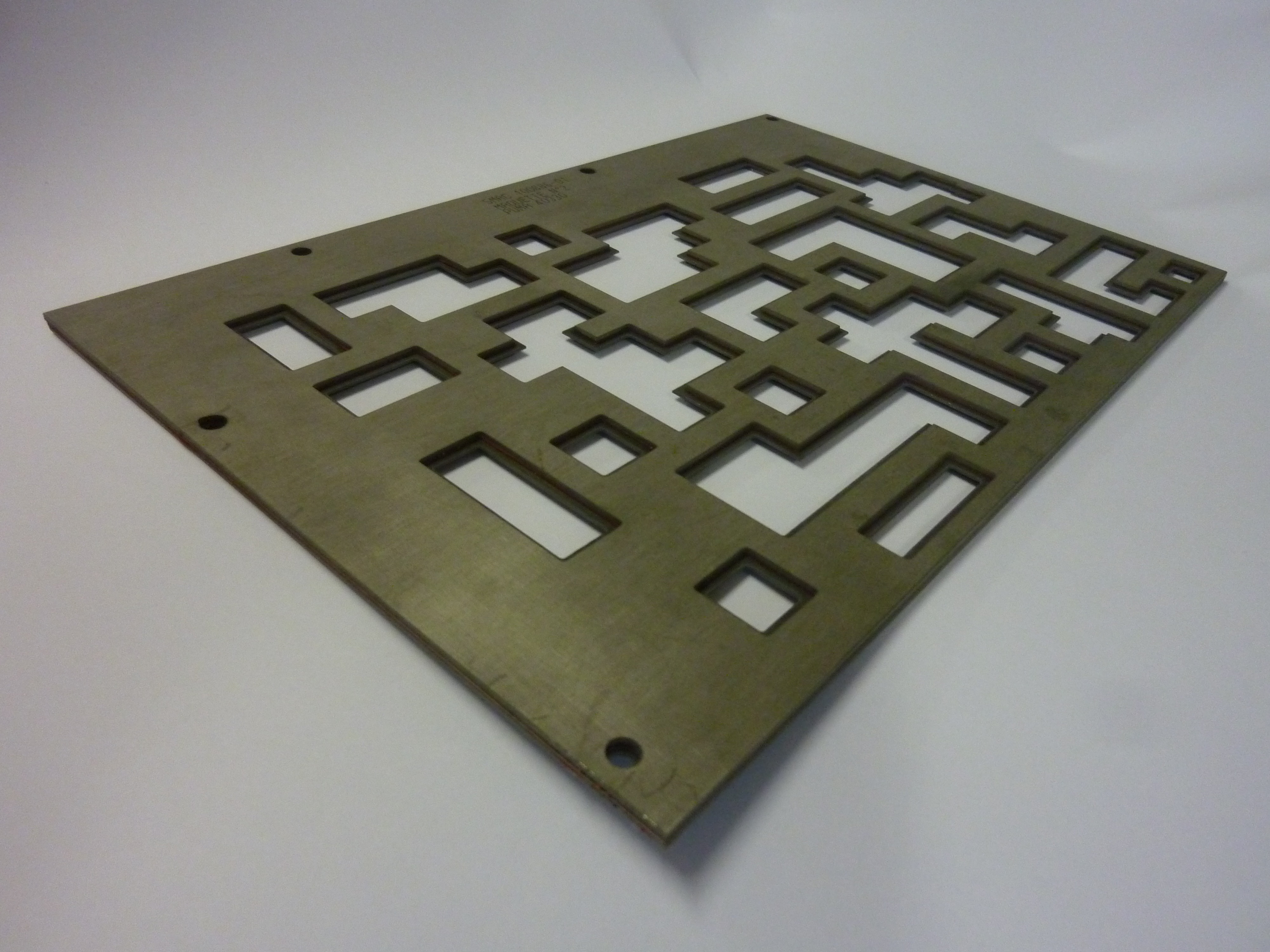} &
\hspace{-2cm}\includegraphics[height=4.cm]{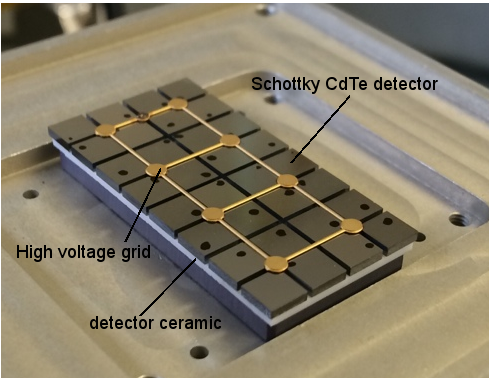}\\
\end{tabular}
\end{center}
\caption[]{\label{fig:design} Top left: View of the coded mask camera
  ECLAIRs. The different sub-systems are shown, except the data processing
  unit in charge of the GRB detection and localization. Top right: View of the
  ECLAIRs detection plane (DPIX) paved with $80\times 80$ Schottky CdTe
  detectors arranged into $8\times 4$ detector matrices called XRDPIX. The
  detection plane is segmented into 8 sectors of 25 XRDPIX. Each sector is
  connected to an electronic chain (ELS) that will work in the photon counting
  mode i.e. each event will be time-tagged with a time resolution of
  $10~\mu\mathrm{s}$ and calibrated in energy in real time. The ELS will also
  perform pattern recognition flagging the events as single or multiple
  events. Bottom left: photo of a mock-up of the coded mask made of a sandwich
  of Ti-Ta-Ti. Bottom right: photo of an XRDPIX.}
\end{figure}

The ECLAIRs GRB trigger camera [\citenum{Triou09}] is based on the technology
of coded mask cameras (e.g. [\citenum{Caroli}]). ECLAIRs is developed by a
consortium of French institutes (PI: IRAP -- CEA-Saclay, IAP \& APC) under the
supervision of the French Space Agency (CNES). Fig~\ref{fig:design} shows the
ECLAIRS sub-systems. Given the capability of the Chinese platform in term of
mass and power available for the science payload, the design of the ECLAIRs
instrument including the Data Process Unit should be compliant to the
following mass and power allocation: 86.5 kg (with a 15\% margin) and 84 W at
most when working in the nominal configuration.

SVOM being a mini-satellite, only $\sim 1000$ cm$^2$ of geometrical surface
could be allocated to the detection plane of ECLAIRs. To increase the
sensitivity in particular to soft GRBs (X-ray Flashes and high redshift GRBs),
it is then mandatory to have a low-energy threshold as low as possible (see
Section~\ref{sec:sensitivity}). The goal is to have a scientific threshold of
4 keV offering an effective area at this energy of $\sim 150$ cm$^{2}$. In
order to reach this goal, we consider the use of $4\times 4$ mm$^2$ and
1\,mm-thickness Schottky CdTe detectors which thanks to their Schottky anode
and high resistivity offer low level of leakage current. The CdTe detectors
will be operated below $-18^\circ\mathrm{C}$ (see
Section~\ref{sec:calibration}). These detectors were provided by Acrorad Co.,
Ltd (Japan). Note that ohmic CdTe and CZT detectors were already used in space
instrumentation (e.g. Swift/BAT [\citenum{barthelmy}] and INTEGRAL/ISGRI
[\citenum{Lebrun}]). However, the low-energy threshold of these instruments is
larger than 10 keV. The flight detectors will be arranged over $8\times 4$
detector matrices coupled with the CEA ASIC IDeF-X [\citenum{Gevin09, Gevin06,
    Gevin}], the so-called XRDPIX (see Fig.~\ref{fig:design};
[\citenum{L13}]), that will pave the detection plane (DPIX) making an array of
$80\times 80$ CdTe pixels mounted on a cold plate in AlBeMet. The IDeF-X ASIC
shows a low intrinsic noise with an Equivalent Noise Charge floor of 33
electron r.m.s. for a peaking time of $6~\mu\mathrm{s}$, a low consumption ($∼
3$ mW/channel) and is radiation hard (Single Event Latchup Linear Energy
Transfer threshold of $56~\mathrm{MeV~cm}^2~\mathrm{mg}^{-1}$). Given the 1-mm
thickness of the detectors and the way the ASIC is working, the high energy
threshold will be set at 150 keV. Therefore, the total energy range is
expected to be from 4 keV to 150 keV. The detection plane will be segmented
into 8 sectors consisting of 25 XRDPIX (i.e. 800 CdTe detectors) to prevent
any failure of the whole plane. Each sector will be connected to an electronic
chain (ELS) that will work in the photon counting mode i.e. each event will be
time-tagged with a time resolution of $10~\mu\mathrm{s}$ and calibrated in
energy in real time thanks to a FPGA. The ELS will also perform pattern
recognition flagging the events as single or multiple events. The electronic
chain is designed to support an event flow from an extremely bright GRB up to
$10^5$ events s$^{-1}$ over the all detection plane with a dead time of less
than 5\%.

The calibrated events will be transmitted in real time via a space wire
connection to the data process unit (UGTS -- [\citenum{Provost13}]) that is in
charge of detecting GRBs as well as managing the DPIX modes and instrument
housekeeping. The UGTS will use the recorded events to reconstruct sky images
every 20\,s to search for the appearance of new sources within the ECLAIRs FoV
by comparison with an on-board catalog of known X-ray sources. The UGTS will
use both a count rate trigger algorithm on different timescales as well as an
image trigger algorithm [\citenum{Schanne13}]. In case of a rate trigger, an
automatic slew of the plate-form will only happen if it is confirmed by the
appearance of a new source in the reconstructed images.

To maximize the GRB detection rate, a large FoV is required in order to be
able to scan a large portion of the sky. Given the space limitations on the
SVOM plate-form, ECLAIRs will have a large FoV of $89\times 89~\mathrm{deg}^2$;
which corresponds to $\sim 2$ steradian. The totally coded FoV is $22.1\times
22.1~\mathrm{deg}^2$ (i.e. $\sim 0.15~{\rm sr}$). A graded passive shield
combining Pb (0.9 mm), Cu (0.1 mm) and Al (0.5 mm) will enclose the detection
plane (DPIX) and the coded mask located at a distance of 46 cm from the DPIX
(see Fig.~\ref{fig:design}). The choice of the material for the graded shield
is a trade-off between a maximum reduction of the cosmic X-ray diffuse
background in the 4-70 keV imaging band [\citenum{Godet09}], mass budget
considerations and the necessity to keep a few instrumental X-ray lines at
higher energies for calibration purposes.
                                                             
The mask will have a transparency of 40\% (the mask transparency was initially
of 30\% and we changed it to 40\% in order to improve the detection of short
GRBs) and the blocking elements in the 4-70 keV band will be arranged into a
quasi-random way. Given that the low energy threshold is expected to be at 4
keV, the use of a material supporting the blocking elements of the mask is not
possible because any material at this energy will absorb a too important
fraction of the incident photons, and hence this will reduce the effective
area at low energy. Instead, the mask should be able to stand alone without
any isolated blocking element. To do so, it will be pre-constrained before
installed on the mechanical structure supporting the passive shield. Because
of this, a special care has been made in the choice of the current baseline
mask pattern to minimize the thermo-mechanical deformation of the mask (see
also Section~\ref{sec:localisation}). The mask will be made of Ta (0.6 mm in
thickness) sandwiched by two Ti sheets of 1 mm in thickness each in order to
ensure the flatness of the mask (see Fig.~\ref{fig:design}). The mask will be
covered by a multi-layer thermal coating insulation (MLI) to prevent optical
photon loading on the detection plane.

\section{Characterizing the performance of the XRDPIX} 
\label{sec:calibration}

The key science driver for the design of the detection plane of ECLAIRs is the
low-energy threshold as mentioned in Section~\ref{sec:design} (see also
Section~\ref{sec:perfo}). In order to ensure that a low-energy threshold of 4
keV could be reached, we defined a detailed test plan to validate the
performance of each sub-system of the XRDPIX from the single CdTe detector to
the complete XRDPIX by performing current, spectroscopic and noise
measurements [e.g. \citenum{R10, L13} -- see also Nasser et al.: paper
  9144-150 this conference]. Several test facilities and software tools were
then developed at the IRAP in order to perform these measurements (see Pons
et al. -- paper 9144-196 this conference).

\subsection{Selection of the Schottky CdTe detectors for the flight model}

\begin{figure}
\begin{center}
\begin{tabular}{c}
\includegraphics[height=6.cm]{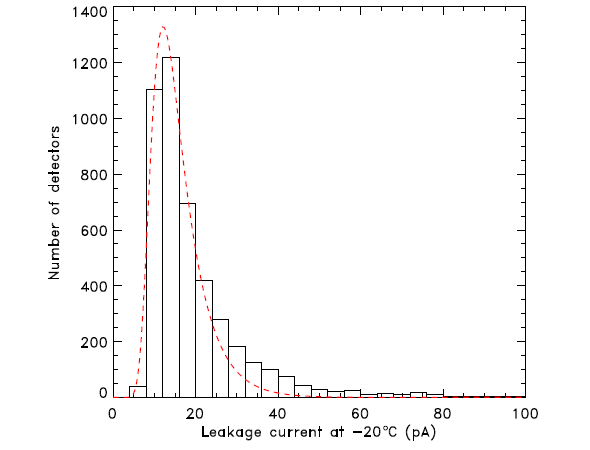}
\end{tabular}
\end{center}
\caption[]{\label{fig:current} Distribution of the leakage current of the
  selected-to-the-flight CdTe detectors as measured at a temperature of
  $-20^\circ\mathrm{C}$ and a bias of -600V. The figure comes from Remou\'e et
  al. (2010) [\citenum{R10}]. The median value is around 12 pA.}
\end{figure} 

An intensive work of leakage current and spectroscopic measurements on the all
population of CdTe detectors purchased ($12288$ detectors) was realized at the
IRAP in order to characterize the properties of Schottky CdTe detectors and to
define selection criteria [\citenum{R10}] that were used to select the 8000
flight-model (FM) detectors (i.e. 6400 detectors for the detection plane and
1600 detectors for spare modules). All the measurements were performed
applying a reverse bias of -600V to the detectors and at two temperatures (the
in-flight temperature of $-20^\circ\mathrm{C}$ and the ambient temperature of
$+25^\circ\mathrm{C}$) in thermal controlled chambers. Note that the current
measurements were made 30\,s after biasing the
detectors. Fig.~\ref{fig:current} shows the distribution of leakage current
measured at $-20^\circ\mathrm{C}$ for selected-to-the-flight detectors. The
median value is around 12 pA. This value is well below the maximum value of
the leakage current (150 pA -- [\citenum{R10}]) required to reach a low energy
threshold of 4 keV at the $6~\sigma$ level considering the total ASIC input
capacitance of 10.5 pF (worst case) and a peaking time of $6~\mu\mathrm{s}$.

The CdTe detectors considered for the detection plane being of Schottky type,
these detectors are known to be unstable with time i.e. their leakage current
increases over time due to the lowering of the Schottky potential height due
to an accumulation of holes on the Schottky contact. We therefore performed
current measurements over time ($\sim 2~\mathrm{h}$) on a sample of more than
6000 detectors in order to characterize the temporal evolution of the leakage
currents. We find that the increase in the leakage current over timescale of a
day should be relatively moderate and this should not be considered as an
issue regarding the ECLAIRs performance, in particular the low energy
threshold [\citenum{R10}]. 

\begin{figure}
\begin{center}
\begin{tabular}{c}
\includegraphics[height=6cm]{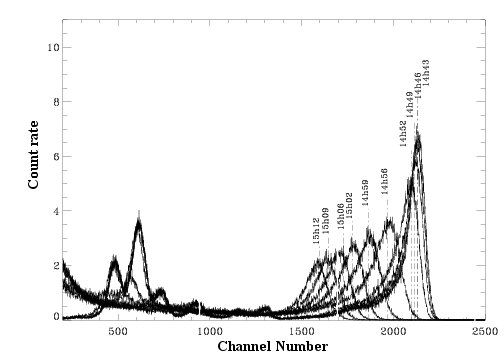}
\end{tabular}
\end{center}
\caption[]{\label{fig:pola} Degradation of the spectroscopic performance of a
  Schottky CdTe detector with time due to the polarization phenomenon. The
  spectrum was measured at $+25^\circ\mathrm{C}$ using a $^{241}$ radioactive
  source. The detector was biased at -100V.}
\end{figure}

The lowering of the Schottky potential height also induces a degradation of
the spectroscopic performance of the detector with time. This is known as the
polarization phenomenon [\citenum{Toyama, cola}]. This degradation is
characterized by a decrease in quantum efficiency (QE), an increase in
spectral resolution (FWHM) and a shift of the photo-peak towards lower
energies with time (see Fig.~\ref{fig:pola}). This phenomenon is enhanced with
the increase in temperature and the lowering of the reverse bias
[\citenum{Toyama, cola, R10}]. It starts immediately after the detectors have
been biased. Initially the degradation in QE and FWHM and the shift of the
photo-peaks are relatively small [\citenum{R10}; Nasser et al.: paper 9144-150
  this conference]. However, after the so-called polarization time ($t_{\rm
  pola}$), the spectroscopic degradation happens more rapidly. [\citenum{R10}]
measured a shift velocity of the $^{241}$Am 59.6 keV line of $V_{\rm shift}
\sim -64$ eV/min before $t_{\rm pola}$ and a polarization time of $t_{\rm
  pola} \sim 11$ min for a bias of -100\,V and a temperature of $+25^\circ{\rm
  C}$. Note that $t_{\rm pola}$ and $V_{\rm shift}$ increase rapidly with the
temperature and the reverse bias. The regular depolarization of the Schottky
CdTe detectors allows to restore the spectroscopic performance of the
detectors. In orbit, it is expected to depolarize the detectors during deep
passages of SVOM across the SAA (the duration of the longer passage outside
the SAA is expected to be around 15 hours).

$\sim 76\%$ of the all population of CdTe detectors purchased passed the
selection criteria.

\subsection{Performance of the XRDPIX}

We focus here on our effort in characterizing in detail the performance of 5
proto-flight XRDPIX modules thanks to dedicated lab measurements that enable
us to explore the behavior of these modules for various parameters (e.g. the
reverse bias HT, the ASIC peaking time $t_{\rm peak}$, the frequency of the
FPGA clock). The main objectives of this work are to: i) improve our
understanding of the physics behind Schottky type detectors (Nasser et al. --
paper 9144-150 this conference; see also [\citenum{R10}]); ii) study how the
XRDPIX performance (e.g. the low-energy threshold, the quantum efficiency, the
gain linearity, the polarization effect, etc.) evolve with these tunable
parameters in order to establish a set of parameters to be used in-flight 
that will guarantee the stability of the instrument performance once in
orbit. We also performed polarization measurements on several XRDPIX in order
to ensure that for the in-flight configuration (see above) the impact of the
polarization phenomenon on the detector spectroscopic performance could be
almost neglected [see Nasser et al.: paper 9144-150 this conference]. A
dedicated thermal-vacuum facility was built at the IRAP in order to realize
all this work. The electronic chain used to perform these tests has the same
functionalities as the in-flight chain. Most measurements were done in the
photon counting mode and at the in-flight temperature of $-20^\circ\mathrm{C}$
using a $^{241}$Am radioactive source.

\begin{figure}
\begin{center}
\begin{tabular}{cc}
\hspace{-1cm}\includegraphics[height=6cm]{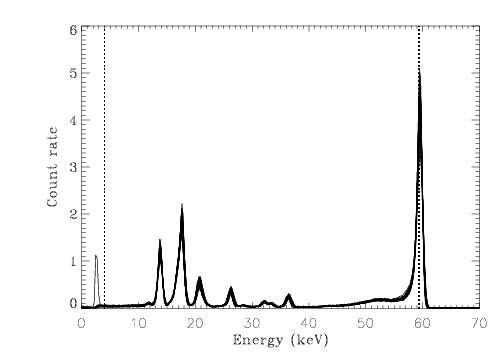} &\hspace{-0.8cm}
\includegraphics[height=7.cm]{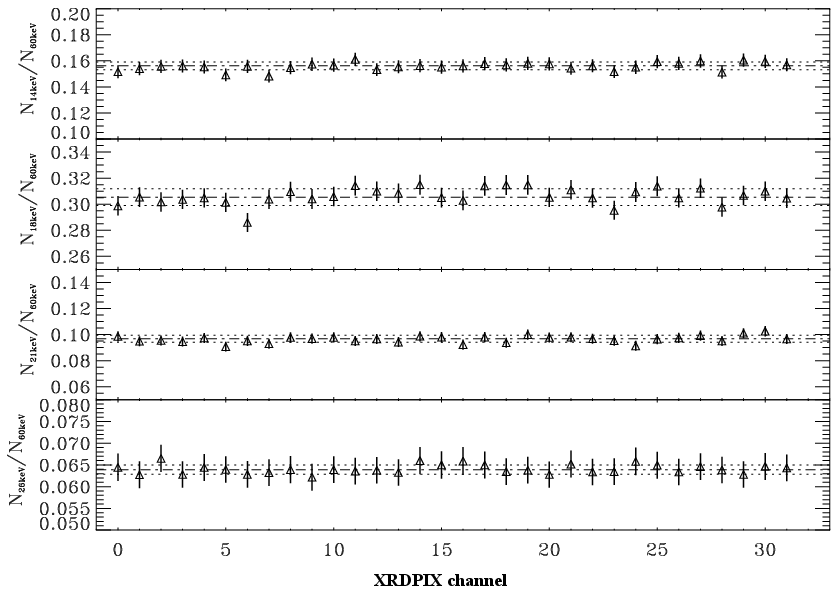} \\
\end{tabular}
\end{center}
\caption[]{\label{fig:result} Left: Superposition of 32 spectra measured on
  one proto-flight XRDPIX (i.e. a 32 CdTe pixel matrix) obtained using a
  $^{241}$Am radioactive source. The measurement was performed in the IRAP
  thermal-vacuum chamber at $T = -20^\circ$C, a bias of -400\,V and a peaking
  time of $4.4~\mu\mathrm{s}$. Note that when the electronic noise peak is
  visible it is below 4 keV (the vertical dotted line). Right: Ratios of the
  number of counts under each $^{241}$Am low energy lines over the number of
  counts under the $^{241}$Am 59.6 keV line for 32 CdTe detectors of one
  XRDPIX. The measurements were also performed in the IRAP thermal-vacuum
  chamber at $T = -20^\circ$C, a bias of -400\,V and a peaking time of
  $4.4~\mu\mathrm{s}$. The errors on the ratio-values are given at the
  $3~\sigma$ level. Note the very good homogeneity of the detectors.}
\end{figure}

We explore a large parameter space with 5 HT-values (from -250 to -450\,V) and
7 $t_{\rm peak}$-values (from 0.95 to 6.8 $\mu\mathrm{s}$). We also considered
the influence of adding an artificial dead-time of $34~\mu\mathrm{s}$ at the
end of the ASIC event reading phase on the stability of the XRDPIX
performance over time. The FPGA clock frequency was set at 4 MHz for all
measurements. This work is still on-going. 

\begin{figure}
\begin{center}
\begin{tabular}{c}
\includegraphics[height=6.cm]{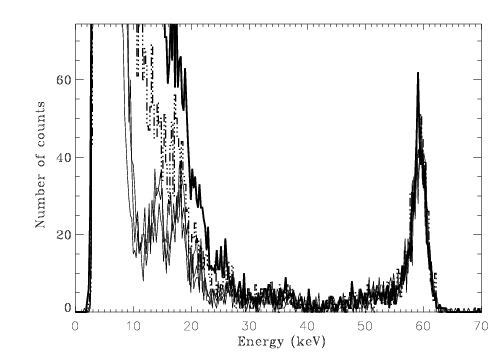}
\end{tabular}
\end{center}
\caption[]{\label{fig:temp} Superposition of $^{241}$Am spectra measured over
  successive epochs. The exposure time for each spectrum is 3 min. The
  measurements were performed at $T=-20^\circ\mathrm{C}$, a bias of -250\,V
  and a peaking time of $6.8~\mu\mathrm{s}$. The thick dotted-dashed spectrum
  corresponds to the first measured spectrum, while the thick solid spectrum
  corresponds to the last measured spectrum. Note the evolution of the ``noise
  peak'' with time, while the spectral resolution of the $^{241}$Am 59.6 keV
  line does not change significantly.}
\end{figure}

We find from preliminary analysis of this work that the overall spectroscopic
response of the CdTe detectors from different XRDPIX appears to be homogeneous
within the different configurations considered as illustrated in
Fig.~\ref{fig:result} (see also Nasser et al. 2014 -- paper 9144-150 this
conference). The absence of the noise peak above 4 keV in most spectra (see
the left panel in Fig.~\ref{fig:result}) indicates the very good performance
of these Schottky CdTe detectors. During the advancement of this work, we also
noticed that \underline{in rare cases} a prominent ``noise peak'' could appear
in some spectra (see Fig.~\ref{fig:temp}). Note that the apparition of this
prominent ``noise peak'' does not degrade too much the FWHM of the $^{241}$Am
59.6 keV line. Thus, we stress that this noise peak is unlikely to be related
to a large increase in the level of the leakage current. Moreover, it appears
that this ``noise peak'' varies with time. So, we suggest that this prominent
``noise peak'' may be due to the development of perturbations maybe related to
the way the electronic chain works. Note that the electronic chain has not
been fully validated yet. We are currently investigating the nature of these
instabilities. We notice that the strength of this parasitic ``noise peak''
could be mitigated by decreasing the ASIC peaking time or by adding an
artificial dead-time of $34~\mu\mathrm{s}$ at the end of the ASIC event
reading phase.

\subsection{Performance of the electronic chain}

\begin{figure}
\begin{center}
\begin{tabular}{c}
\includegraphics[height=6.5cm]{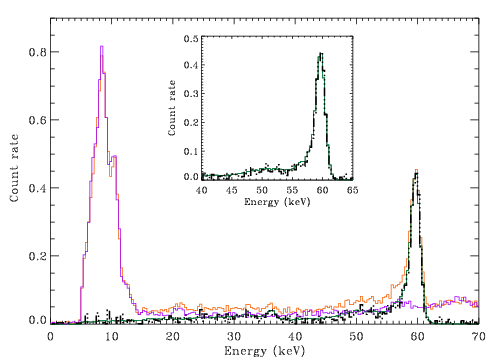}
\end{tabular}
\end{center}
\caption[]{\label{fig:tandem} Spectra of one XRDPIX pixel as measured at the
  Tandem particle accelerator facility in order to investigate the impact of
  proton irradiation on the photon energy coding. The detectors were polarized
  at -400\,V and we used a peaking time of $4.4\mu\mathrm{s}$. Green: the
  $^{241}$Am spectrum with the proton beam off (REF). Only the $^{241}$Am 59.6
  keV line is visible, because the other low energy lines were absorbed by the
  front entrance window of the thermal-vacuum chamber used during this
  experiment; Orange: the $^{241}$Am spectrum with the proton beam on (ALL);
  Blue: the spectrum induced by protons only (DARK). The thick dashed spectrum
  (RESTORE) corresponds to the contribution of the photons in the ALL spectrum
  (i.e. RESTORE $=$ ALL $-$ DARK). The close-up shows a comparison of the
  reference $^{241}$Am spectrum (REF) and the RESTORE spectrum. We corrected
  the RESTORE spectrum for the effect of dead time (see the text). Note that
  the $^{241}$Am 59.6 keV line is not perturbed by the proton irradiation. The
  lines at low energies seen in the ALL and DARK spectra are fluorescence
  lines from the experimental setup materials induced by the proton
  irradiation.}
\end{figure}

In addition to the measurements done at the IRAP, we also performed two test
campaigns in 2010 and 2012 using the particle accelerator facility
Tandem\,\footnote{http://ipnweb.in2p3.fr/tandem-alto/tandem/} at the IPN Orsay
in France. The goal was to simulate the in-flight particle radiation
environment of SVOM along its LEO orbit outside the SAA in order to study the
impact on the ECLAIRs electronic chain: estimate the desaturation time and the
impact on the energy coding of the photons [\citenum{Nasser14}]. To do so, one
XRDPIX operated under flight conditions (vacuum and $-20^\circ\mathrm{C}$)
was irradiated with an incident flux of high energy (20 MeV) protons and a
$^{241}\mathrm{Am}$ source.  As illustrated in Fig.~\ref{fig:tandem} for one
XRDPIX pixel, the spectral response of the XRDPIX to photons is not degraded
once it is irradiated with protons. The only difference from the spectra
measured with and without the beam of protons on is the loss of some photon
events due to the increase of the dead time in the former case. The level of
dead time is estimated to be around 8.5\%. This experiment shows the
robustness of the electronic chain to the particle flux irradiation
expected on the SVOM orbit [\citenum{Nasser14}].

\section{Conclusion} 
\label{sec:conclusion}

With a launch expected around 2020, SVOM will be one of the main GRB missions
(maybe the only one) combining regular alerts to the world community, a
multi-wavelength coverage and rapid slewing capability. In addition, SVOM will
work in a very exciting instrumental context with new ground and space
facilities being operational such as the James Webb Space Telescope (mid and
near IR) and the European-Extremely Large Telescope (near-IR/optical), Atacama
Large Millimeter Array (sub-mm, mm), LSST (optical), SKA-pathfinder (radio)
and Cherenkov Telescope Array at very high energy (TeV) on the ground. Such
facilities with their very deep limiting sensitivity and observing strategy
will open the era of multi-wavelength synoptic astronomy. This will be
beneficial for studying the GRB afterglow emission over the broadband
electromagnetic spectrum as well the environment of GRBs, their host galaxies
in particular at high redshift. SVOM will also provide short GRB triggers to
GW facilities such as aLIGO and aVirgo in order to search for GW signals from
neutron star mergers.  Synergy between SVOM and all these photonic and
non-photonic facilities will enable to fulfill some interesting studies in
cosmology and fundamental physics for GRB and non-GRB science.

In this paper, we presented the GRB trigger camera ECLAIRs on-board SVOM. We
showed that the instrument performance will enable SVOM to detect a sample of
at least 200 GRBs over the 3 yr mission lifetime, including a sample of a few
high-z GRBs and low-z sub-luminous GRBs. Fast distribution of the SVOM alerts
(from ECLAIRs, MXT, VT \& GFTs) to the world community will enable to
coordinate follow-up observations with ground and space facilities enabling
GRB-based cosmological and fundamental physics studies. The performance of
ECLAIRs are also expected to be adequate to observe a handful sample of other
energetic transient events (e.g. TDEs, supernovae) that will also be observed
at other wavelengths thanks to the MXT, VT and the ground instruments. We also
stressed that ECLAIRs thanks to its sensitivity at low energy and its wide FoV
will help following up transient events observed by other facilities in
different wavelengths. We also presented the detailed characterization plan to
ensure that the science requirements of the instrument could be reached, in
particular the low energy threshold of 4 keV. From our lab measurements on
characterizing the XRDPIX performance, we find that the preliminary
spectroscopic results are very promising.


\bibliography{report}   
\bibliographystyle{spiebib}   

\end{document}